# A Parallel Processing Algorithm for Computing Short-Range Particle Forces with Inhomogeneous Particle Distributions

ROBERT C. FERRELL

*Thinking Machines Corporation, 254 First Street, Cambridge, MA 02142*

*e-mail: robert@ferrell.com*

EDMUND BERTSCHINGER

*Department of Physics, MIT Room 6-207, Cambridge, MA 02139*

*e-mail: edbert@mit.edu*

## Abstract

We present a computational algorithm for computing short range forces between particles. The algorithm has two distinguishing features. First, it is optimized for multi-processor computers, and will use as many processors as are available. Second, it is optimized for inhomogeneous, dynamic particle distributions; for any distribution the computational load is distributed evenly to all processors, and the communication time is less than 15% of the total run time.

In this talk we present our new algorithm. We developed the program for a grand-challenge problem in cosmology, simulation of the formation of large-scale structure in the universe. This simulation, run on the Thinking Machines Corporation CM-5, uses the particle-particle/particle-mesh (PPPM) [Hockney and Eastwood 1988] algorithm. The particle-particle phase is computed using the algorithm we describe in this paper. We discuss this and other applications.

## 1 Introduction to the Problem

Particle methods form a class of numerical techniques. One of the advantages of particle techniques is that particles can be concentrated in physically interesting regions. Thus, for a fixed number of degrees of freedom, it is possible to increase resolution in those regions. Another ad-vantage is that for certain problems the particle dynamics are known, but no field theory is available. Typical uses of particle methods are shown in figures 1, 2 and 3.

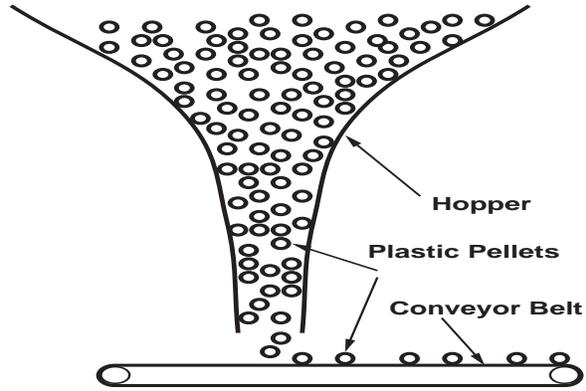

Figure 1: *Plastic pellets flowing through a hopper.*

Figure 1 shows the flow of particles of plastic through a piece of machinery. In this example the use of particles is physically motivated. The force is a contact force- pairs of particles have a force between them if they are touching.

Figure 2 shows fuel injection droplets in a simulation of a combustion engine. The particles represent collections of fuel droplets. These droplets collide, coalesce and break-up. The "force" between the droplets is a pair-wise statistical process between nearby particles.

Figure 3 shows mass particles in a simulation of formation of structure in the universe. The particles each represent a mass of about 10% of a galaxy. The particles are strictly numerical, not physical, particles. The force between particles is Newtonian gravity. Although gravity is infinite-ranged, the PPPM method [Hockney and Eastwood 1988] separates the force on each prticle into a long-range component computed with a mesh and a short-range component for sub-mesh scales. We have described our parallel implementation of the particle-mesh algorithm elsewhere. [Ferrell and Bertschinger 1994] The short-range component is our concern here.

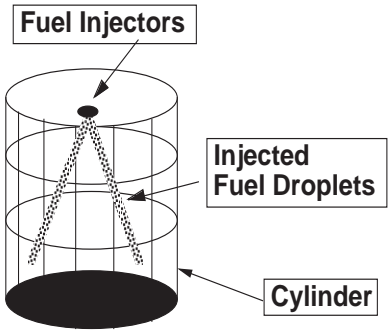

Figure 2: *Fuel droplets in an engine cylinder.*

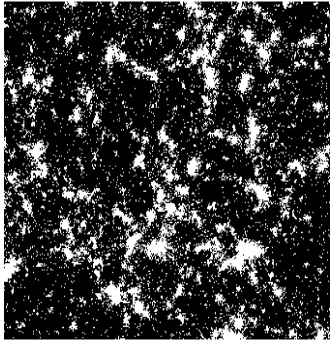

Figure 3: *Simulation of formation of structure in the universe.*

All three of these examples share the properties:

1. The force between the particles has a finite range and vanishes beyond some $R_{cut}$, which is much smaller than the length scale of the simulation volume.

2. The distribution of particles changes with time, perhaps significantly.

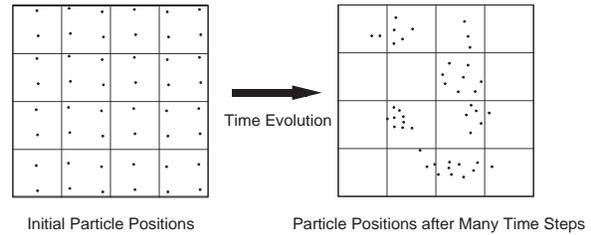

Figure 4: *Inhomogeneous particle distribution yields poor load balancing if a naive algorithm is used. Each square represents a processor.*

3. Particles that start near each other may end up far apart, and vice-versa.

4. The density of particles is non-uniform and time dependent — some particles have many neighbors, some few or none.

For any computational system this presents challenges, and on a multi-processor computer these challengers are particularly acute. In particular:

1. Any distribution of particles to processors will have to evolve as the simulation progresses.

2. Distributing the work to the processors is non-trivial, since any even division of spatial volume or particle number will yield uneven processor utilization.

Figure 4 shows this clearly. Assigning particles to processors based simply on spatial location assigns widely varying numbers of particles to different processors. Assigning the same number to each processor will yield widely different amounts of work and communication between the processors to compute short-range forces.

A lesson we have learned is that the most important issue in parallelism is to organize the data for efficient computation. This means it is desirable to do some bookkeeping and communication at the beginning of each time step which will allow the computation to proceed at full speed.

To this end, our algorithm divides into two phases. In the first phase we organize the particle data into a good data structure for computation. In the second phase we perform the force computation, based on the data struc-ture of part one. The key to our success was finding the right data structure to allow for efficient computation.

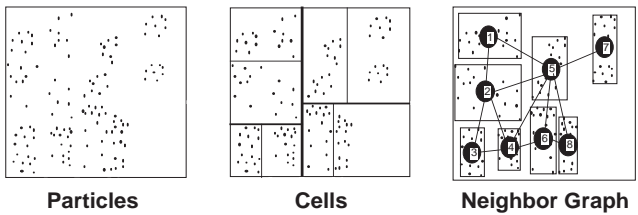

Figure 5: *Divide the particles into cells, each containing CellSize particles, then connect cells to their neighbors.*

The first part of this paper introduces the fundamental concepts of the method. We then present some techniques for optimizing the computation and communication on multi-processor systems. Finally, we present some results from the code we have developed.

## 2 The Neighbor Graph and the Force Calculation

Figure 5 shows the two steps of phase one. First the particles are grouped into "cells". Each cell contains $CellSize$ particles. The cells have varying spatial sizes, depending on the local particle density.

Next, cells which are nearby to each other are identified as neighbors. A graph is constructed, with the cells as nodes in the graph and an edge of the graph between neighboring cells. (Cells do not have to be adjacent to be neighbors, merely within $R_{cut}$ of each other. The exact definition of "neighbor" is given below.) We call this the "cell neighbor graph", or just the "neighbor graph". A cell is a neighbor of itself, although that edge is not drawn.

In phase two the force between pairs of particles is calculated. Only cells which are identified as neighbors have particles within $R_{cut}$ of each other. Therefore, the force computation is reduced from $\mathcal{O}(N^2)$ to $\mathcal{O}(Degree\_of\_Neighbor\_Graph * CellSize^2) \ll N^2$.

### 2.1 Benefits of this Algorithm

An important advantage of this data structure (cells and a neighbor graph ) is that the force computation is simple. Pick an edge of the neighbor graph. Say the edge connects cells $i$ and $j$. Compute the force between all particles in $i$ and all particles in $j$. Next pick another edge, and so on, until all the edges are processed.

The parallelism is obvious — we distribute the edges of the neighbor graph over the processors. The order in which edges are computed does not matter, so all processors can run simultaneously. [1] We will see later that the order of processing the edges can be optimized to minimize communication costs.

Each time an edge is computed, the position and the force data for all the particles in both cells connected by that edge has to be communicated to and from the processor computing on that edge. This communication may be a cache-load or it could be data communication from one physical processor to another. The relative cost of this communication can be kept low. The communication is $\mathcal{O}(CellSize)$ and the computation is $\mathcal{O}(CellSize^2)$. Therefore, no matter what the rates are for the communication and computation, $CellSize$ can be chosen large enough that the communication cost is lower than the computation cost. (If $CellSize$ is too large, the cells become larger than $R_{cut}$ and the fraction of wasted force calculations — calculations on particle pairs separated by more than $R_{cut}$ — increases. Hence the efficiency decreases if $CellSize$ is too large. We discuss this issue further later.)

To summarize, the steps of an efficient parallel algorithm are:

1. Construct the data structure for computation.
   (a) Organize the particles into cells.
   (b) Build a neighbor graph of the cells.
2. Distribute the edges of the neighbor graph to the processors.

### 2.2 Comparison with Other Work

We emphasize that this method is new and distinct from methods which divide the problem spatially, *e.g.* [Lomdahl *et al.* 1993]. It is also distinct from methods which distribute the particles directly [Theuns 1994]. [Plimpton and Heffelfinger 1992] distributes pair forces to processors, but do not use cells. This limits the number of particles which can be simulated. For a review of prior work done on parallel algorithms for molecular dynamics see [Beazley *et al.* 1995].

---

[1] Floating point arithmetic is not commutative, so the result depends slightly on the order of computation. We are assuming that the simulation is stable against numerical round-off of this sort, and consequently ignore this small effect.

Our method is the first efficient, scalable, load-balanced algorithm we know for computing arbitrary short-range two-body forces. Most parallel algorithms are optimized for nearly homogeneous particle distributions, such as arise in solid crystal simulations. For force laws with simple multipole expansions (such as gravity) tree algorithms [Barnes and Hut 86] have been developed. [Warren and Salmon 94] pioneered efficient parallel tree algorithms which, for strongly clustered distributions, have the potential to be more efficient than PPPM. However, our method is more general since it can be applied to any pair force, not only forces with simple multipole expansions. Further, PPPM can be made more efficient using adaptive mesh refinement [Couchman 91, Bertschinger and Gelb 91]. Now that we have parallelized both PP and PM in PPPM we can implement adaptive mesh refinement.

## 3 Organizing the Particles

In this section we describe how the neighbor graph is constructed. As we noted, this is the first phase of each time step. The keys to efficiency and speed are

1. A fast method for dividing the particles into cells.

2. A fast and load balanced way of constructing the neighbor graph.

In this section we give some details of how we accomplish these two steps.

### 3.1 Organizing the Particles into Cells

Any method we choose must be: 1) Fast, so we can re-organize the particles each time step, and 2) construct cells which are compact, *ie.* particles which are near each other should be put into the same cell.

We have adopted the well known recursive, orthogonal bisection (R.O.B.) [Cormen *et al.* 1990] as a good compromise. The choice of decomposition method is not central to the rest of the algorithm, so long as it is fast. For particular problems we may want to choose something other than R.O.B..

Figure 6 shows how R.O.B. works. For $NPart$ particles and a cell size of $CellSize$ there are $\log_2(NPart/CellSize)$ iterations. This results in $NCell = NPart/CellSize$ cells. (Note that we require that $NCell$ be a power of 2.) At each iteration, each cell is subdivided into two equal occupation volumes. The

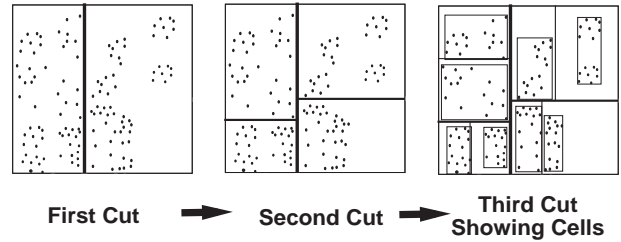

First Cut → Second Cut → Third Cut Showing Cells

Figure 6: *At each iteration, each volume is divided into two parts, each with the same number of particles. The splitter dimension is chosen to divide the dimension with the greatest spread between particles.*

splitters are planes in three dimensions. For each cell the splitter plane is chosen to cut the dimension with the greatest spread.

Recursive Orthogonal Bisection is fast on the CM-5 because the sorting routine is fast, and all cells can be split in parallel. The data structure for the particle positions is an array $X(NPart, 3)$. This array is organized so that particles in the same cell are in contiguous segments. The CM-5 sort routine can sort all segments, independently and simultaneously.

(We gave some thought to using an adaptive scheme, so that the cells need not be re-constructed each time step. Again, for certain problems this may be appropriate. For our purposes, the run time required to organize the particles is less than 10% of the total time. This, combined with the fact that R.O.B. is so easy to code, made us decide that the complexity of an adaptive scheme was not worth the small gain in performance.)

### 3.2 Constructing the Neighbor Graph

For this step, the task is to identify all pairs of cells which have particles less than $R_{cut}$ apart. This must be done quickly, with particular care to avoid load imbalance.

The data structure for the neighbor graph is NGraph($Degree\_of\_Neighbor\_Graph$, 2). This is a list of $Degree\_of\_Neighbor\_Graph$ 2-vectors. Each 2-vector $(i, j)$ represents an edge in the neighbor graph , connecting cell $i$ with cell $j$. If edge $(i, j)$ is in the neighbor graph then $i$ and $j$ are neighbors, otherwise $i$ and $j$ are not neighbors. "Neighbor" is defined below.

(Notice that we could have stored the neighbor graph as an $NCell \times NCell$ matrix. A 1 in element $(i, j)$ would indicate that $(i,j)$ are neighbors, a 0 that they are not. However, the neighbor graph is sparse, with less than 1% non-zero entries. Therefore, a neighbor graph matrix would be mostly 0's and highly wasteful of memory. Those familiar with sparse matrix operations will recognize that our neighbor graph data structure is just the standard representation for a sparse matrix.)

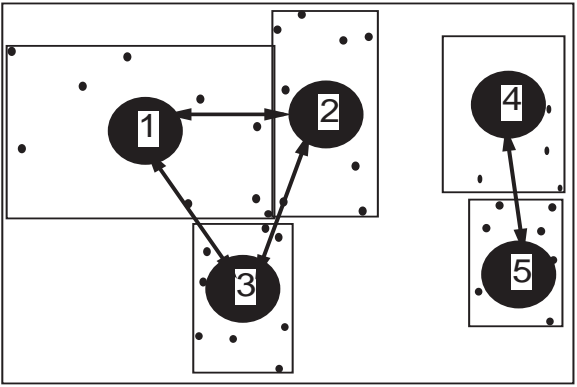

Figure 7: *Cells that overlap, or are within $R_{cut}$ of each other, are identified as neighbors.*

Figure 7 shows the idea behind the construction of the neighbor graph. Cells which overlap, or are within $R_{cut}$, are identified as neighbors More specifically,

$$(i,j) \in \text{Neighbor Graph} \iff \text{ for each } l \in [x,y,z]$$
$$\left\{ \begin{array}{ll} Upper(i)_l & +R_{cut} \geq Lower(j)_l \quad \text{and} \\ Lower(i)_l & \leq Upper(j)_l + R_{cut} \end{array} \right\}$$

Where $(i,j)$ is the edge connecting cell $i$ to cell $j$, $Upper(i)_l$ is the upper spatial bound of cell $i$ along dimension $l$ and $Lower(i)_l$ is the lower spatial bound of cell $i$ along dimension $l$.

With this definition we identify all pairs of cells $(i,j)$ for which there is at least one particle in cell $i$ and one particle in cell $j$ which are less than $R_{cut}$ apart. However, we may also identify as neighbors cells which do not have any particles within $R_{cut}$ of each other. We have found that this definition is a good compromise between speed and efficiency. For highly clustered distributions, typically less than 5% of the edges in the neighbor graph are spurious.

### 3.3 Parallelism and Load Balancing

In the remainder of this section we discuss the important concepts that relate to an efficient multi-processor implementation for constructing the neighbor graph. The the construction be fully parallelized and load balanced. The degree of the nodes of the neighbor graph (that is, the number of neighboring cells of each cell) will vary widely from cell to cell. Therefore, we must take care that our algorithm does not have a serial component which is proportional to the degree of a node.

We need some terminology. Each edge $(i,j)$ has a head and a tail. $i$ is the tail, $j$ is the head, and edge $(i,j)$ points from cell $i$ to cell $j$. An "outgoing edge" from $i$ is an edge with tail $i$. An "incoming edge" to $j$ is an edge with head $j$. The lines in Figure 7 each represent two edges, an incoming edge and an outgoing edge.

The neighbor graph construction has four steps:

1. For each cell, identify the neighbor cells and build an edge for each neighbor (Figure 7).

2. Construct the data structure for the edges and connect the tails of the edges to the appropriate cell (Figure 8).

3. For each edge connect the head of that edge to the neighbor it points to (Figure 9).

4. Prune the neighbor graph to remove redundant edges (Figure 10).

Figure 7 shows the first step: identifying and counting how many cells overlap each other. For instance, cell 1 overlaps 2 other cells, so cell 1 will have 3 edges from it (including the one to itself, which is not shown).

For the next two steps, the edges are constructed and connected to their tail and head. To connect the tails we need to replicate each cell number according to how many neighbors that cell has. Figure 8 shows how cells are replicated. The replication step is fully parallelized. Where a serial algorithm might loop over nodes, and replicate each node, this is not effective for a parallel processor because nodes have many different numbers of neighbors. Instead, we use the segmented scan copy operator [Chatterjee *et al.* 1990, Hillis and Steele 1986], which distributes the work evenly to all processors.

Figure 9 shows the third step, connecting the heads of the edges. At each overlap, all cells which overlap are neighbors. Since each edge is a (tail,head) pair, and we have already constructed the tails, we know that the mapping from outgoing edges to incoming edges will be a 1-to-1 permutation. Figure 9 shows the permutation, which we term a transpose, since for each overlap point it is like a transpose of the tail list. Finding the permutation index and permuting the data are both parallelized, load-balanced, operations.

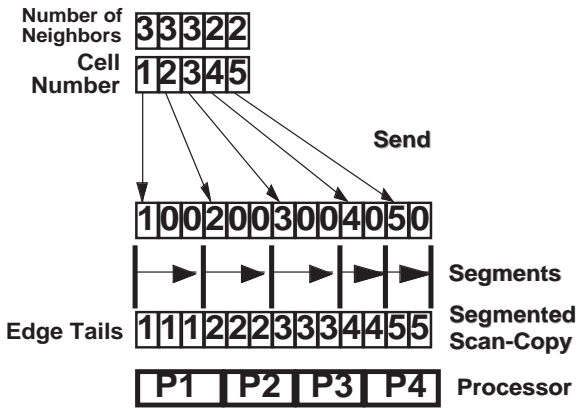

Figure 8: *The node number for each cell is replicated enough times to draw an edge from that cell to each of its neighbors.*

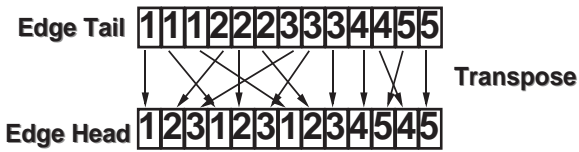

Figure 9: *The permutation that generates the head number from the tail number is like a transpose.*

The final step for construction of the neighbor graph is to remove any redundant edges. Due to Newton's 3rd law, the neighbor graph should contain either $(i,j)$ or $(j,i)$, but not both. We arbitrarily choose that if $i$ and $j$ are neighbors, then $(i,j) \in$ Pruned Neighbor Graph $\iff i \leq j$.

In addition, $(i,j)$ should be contained in the pruned neighbor graph only once. Figure 10 shows how we remove redundant edges by sorting the neighbor graph. First, we sort according to the tail of each edge, so that all outgoing edges for the same cell are adjacent, in the same segment. Next, we sort each segment according to the head of each edge. This organizes identical edges together, and the selection mask picks only a single one.

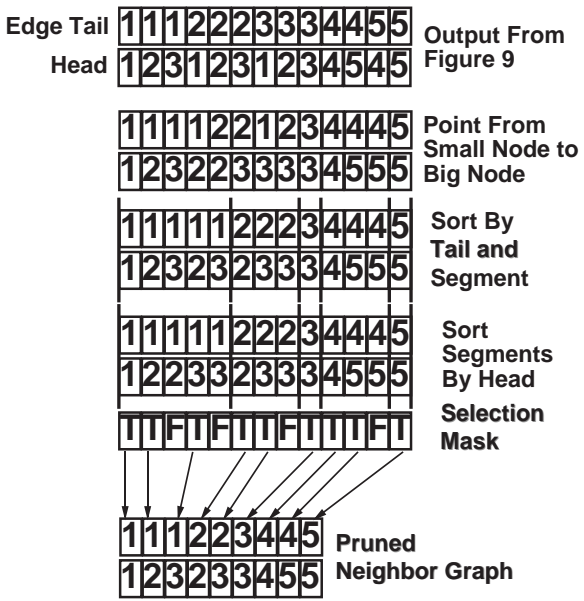

Figure 10: *Redundant edges are pruned from the Neighbor Graph.*

All computation is done using the pruned neighbor graph. In the following discussion, neighbor graph refers to the pruned neighbor graph.

To summarize, in this section we described a parallel, load balanced, algorithm for constructing the neighbor graph. For any distribution of particles the time for construction depends on the total number of edges in the neighbor graph and not on the maximum number of neighbors of any particular cell. In practice, for a heavily clustered particle distribution sorting the particles into cells and constructing the neighbor graph consumes no more than 10% of the run-time.

## 4 Issues for Efficient Computation

As we described in section 2, the effective load balancing of the force computation, for any particle distribution, is due to the fact that the edges of the neighbor graph are distributed among the processor nodes.

The only property of the neighbor graph required for a correct algorithm is that it must contain an edge between every pair of cells which contain particles less than $R_{cut}$ apart. Which edge is distributed to which processor, and which order the edges are computed, are freedoms which can be exploited for optimization.

We organize the neighbor graph, and distribute it to processors, so that each processor loads cell particle position and force data as few times as necessary. In particular, if cell $i$ has many neighbors, we distribute the outgoing edges of cell $i$ to the same processor (or the same few processors, if $i$ has many outgoing edges). Thus, that processor can cache the data for all particles in cell $i$ until all neighbors of $i$ have been computed. In addition, we order the computation so that once a processor has loaded the data for cell $i$ it completes all the computation on that cell (*i.e.* computes all edges whose tails are $i$ ) before proceeding to the next cell.

These are familiar blocking and caching techniques. Although they are usually used only for regular data structures, because of the organization of our algorithm we can apply these techniques to the problem of computation on an irregular data structure. This further emphasizes that the algorithm described here is useful for all multi-processor systems, not merely distributed memory massively parallel processors.

## 5 Conclusion

We implemented a short-range force computation algorithm on the Thinking Machines Corporation CM-5E, a massively parallel processing computer. Our work is distinct from previous work because we optimized the algorithm for inhomogeneous particle distributions. The code we developed is being applied to PPPM simulations to study cluster and structure formation in the universe. For this application the force computation takes approximately 75% of the run-time. The remaining 25% is used for overhead in organizing the data for computation and in communicating data between processors. This is remarkably high performance for a parallel program on a data set with no a priori structure. On a CM-5E the force computation runs at approximately 25 MFlops/Processor. The algorithm is scalable, so the performance (and throughput) increases linearly with the number of processors.

The algorithm is portable to other multi-processor computers and to other applications. In the future we expect to port the code to the IBM SP/2, and perhaps to the SGI PowerChallenge. On either of these systems we expect to increase the efficiency of the computation.

On the CM-5E, due to the nature of the hardware it is necessary to compute all the particle-pair forces for each edge $(i,j)$. In the case where cells $i$ and $j$ are much larger than $R_{cut}$ and the particles are weakly clustered, this can result in computing many forces which vanish. This is the inefficiency that can arise from a large $CellSize$, which we mentioned in section 2.1. Thus there is reason to choose $CellSize$ large enough that the communication is limited to 15% of the run time, but small enough that the number of vanishing forces computed is not great.

For a highly clustered distribution of particles, choosing $CellSize = 64$ yielded an efficiency (number of non-zero forces/total number of forces computed) of 42% for our parallel algorithm. Computing on the same data set with a "standard" serial PP algorithm [Hockney and Eastwood 1988] yielded an efficiency of 59%. This emphasizes that our algorithm is well suited for highly clustered particle distributions — we pay only a small, constant, penalty for using multiple processors and the throughput increases linearly with the number of processors. On the other hand, for weakly clustered particle distributions the efficiency on the CM-5E can be low, for the reasons mentioned above. We anticipate higher efficiencies after we migrate to other systems. On systems with more independent processing nodes, such as the SP/2 or the PowerChallenge, it will be possible to decrease the number of vanishing forces computed. In some cases this will significantly increase the throughput.

## Acknowledgments

The authors would like to thank Dennis Gannon for providing the idea of organizing the particles into cells. This work was supported in part by NSF grant ASC93-181815. We thank the director of NCSA for a discretionary allocation of supercomputer time for code development and testing.

## References

[Barnes and Hut 86] J. Barnes and P. Hut, "A Hierachical $\mathcal{O}(N \log N)$ Force-Calculation Algorithm", *Nature*, **324**, 446 (1986).

[Beazley et al. 1995] D. M. Beazley, P. S. Lomdahl, N. Grønbech-Jensen, R. Giles, P. Tamayo, In preparation.

[Bertschinger and Gelb 91] E. Bertschinger and J. Gelb, "Cosmological N-Body Simulations" *Computers In Physics*, **2**, 164, 1991.

[Chatterjee et al. 1990] S. Chatterjee, G. Blelloch and M. Zhagha, "Scan Primitives for Vector Computers",

in *Proceedings Supercomputing '90*, IEEE Computer Society Press (Los Alamitos, 1990).

[Couchman 91] H. M. P. Couchman, "Mesh-Refined P3M: A Fast Adaptive N-Body Algorithm", *Astrophys. J. (Letters)*, **368**, L23 (1991).

[Cormen *et al.* 1990] T. H. Cormen, C. E. Leiserson and R. L. Rivest, *Introduction to Algorithms*, McGraw-Hill, (New York 1990).

[Ferrell and Bertschinger 1994] R. Ferrell and E. Bertschinger, "Particle-Mesh Methods on the Connection Machine", *Int. Jour. of Mod. Phys. C*, **5**, 933 (1994)

[Giles and Tamayo 1992] R. Giles and P. Tamayo, *Technical Report TR-234*, Thinking Machines Corporation (Cambridge, 1992).

[Hillis and Steele 1986] W. D. Hillis and G. L. Steele, Jr., "Data Parallel Algorithms", *CACM* **29**, 12 (1986).

[Hockney and Eastwood 1988] R. W. Hockney and J. W. Eastwood, *Computer Simulation Using Particles*, Adam Hilger (Bristol, 1988).

[Lomdahl *et al.* 1993] P. S. Lomdahl, D. M. Beazley, P. Tamayo and N. Grøbech-Jensen, "Multi-million Particle Molecular Dynamics on the CM-5." *Int. Jour. of Mod. Phys. C*, **4**, 1074 (1993)

[Plimpton and Heffelfinger 1992] S. Plimpton and G. Heffelfinger, "Scalable Parallel Molecular Dynamics on MIMD Supercomputers", in *Procs. of the High Performance Computing Conference 92*, IEE Computer Society (1992).

[Theuns 1994] T. Theuns, *Comp. Phys. Comm.* **78**, 328 (1994).

[TMC 1993] *CMSSL for CM Fortran, Version 3.1 Beta 2*, Thinking Machines Corporation (Cambridge, 1993).

[Warren and Salmon 94] J. Salmon and M. Warren, "Fast Parallel Tree Codes for Gravitational and Fluid Dynamical N-Body Problems", *Int. J. of Supercomputer App.*, **2**, 129 (1994).